\documentclass[aps,pre,amsmath,amssymb,lengthcheck,showpacs,superscriptaddress]{revtex4-1}
\usepackage{graphicx}
\usepackage{color}
\usepackage{amsfonts}
\usepackage{physics}
\usepackage{braket}

\newcommand{\arctanh}{\operatorname{arctanh}}

\begin{document}


\title{Vibrational spectrum derived from the local mechanical response in disordered solids}

\author{Masanari Shimada}
\email{masanari-shimada444@g.ecc.u-tokyo.ac.jp}
\affiliation{Graduate School of Arts and Sciences, The University of Tokyo, Tokyo 153-8902, Japan}
\author{Hideyuki Mizuno}
\affiliation{Graduate School of Arts and Sciences, The University of Tokyo, Tokyo 153-8902, Japan}
\author{Atsushi Ikeda}
\affiliation{Graduate School of Arts and Sciences, The University of Tokyo, Tokyo 153-8902, Japan}

\date{\today}


\begin{abstract}
The low-frequency vibrations of glasses are markedly different from those of crystals.
These vibrations have recently been categorized into two types: {spatially extended} vibrations, whose vibrational density of states (vDOS) follows a non-Debye quadratic law, and quasilocalized vibrations (QLVs), whose vDOS follows a quartic law.
The former are explained by elasticity theory with quenched disorder and microscopic replica theory as being a consequence of elastic instability, but the origin of the latter is still debated.
Here, we show that the latter can also be directly derived from elasticity theory with quenched disorder.
We find another elastic instability that the theory encompasses but that has been overlooked so far, namely,
the instability of the system against a local dipolar force.
This instability gives rise to an additional contribution to the vDOS, and the spatial structure and energetics of the mode originating from this instability are consistent with those of the QLVs.
Finally, we construct a model in which the additional contribution to the vDOS follows a quartic law.
\end{abstract}


\maketitle

{\it Introduction.} ---Glasses exhibit markedly different low-frequency vibrational properties than crystals do~\cite{Buchenau1984Neutron,Laird1991Localized, Schober1991Localized}.
The low-frequency vibrations of crystals are universally described by the Debye theory, in which plane waves, or phonons, play a central role as elementary excitations~\cite{Kittel1996Introduction}.
The Debye theory, however, cannot be applied to glasses.
Glasses have many more vibrational modes than are predicted by the Debye theory in the low-frequency region.
This manifests as a peak at approximately 1 THz in the vibrational density of states (vDOS) $g(\omega)$ divided by the square of the frequency $\omega^2$, called the boson peak (BP)~\cite{Buchenau1984Neutron}.
Even below 1 THz, glasses have spatially localized vibrations called quasilocalized vibrations (QLVs)~\cite{Laird1991Localized, Schober1991Localized} in addition to phonons.
Since one expects that vibrations of solids will be phonons in the low-frequency region~\cite{Leonforte2005Continuum,Monaco2009Anomalous}, these anomalies are counterintuitive and have attracted considerable interest for several decades.

Recent numerical simulations have revealed quantitative properties of the anomalous vibrations of glasses.
Simulations of weakly coordinated jammed packings have established that the vDOS follows a so-called non-Debye scaling of $g(\omega)\sim\omega^2$ around the BP frequency, independent of the number of spatial dimensions $d$~\cite{Charbonneau2016Universal}.
This scaling, however, breaks down at a certain finite frequency, below which phonons and QLVs coexist down to zero frequency~\cite{Lerner2016Statistics,Mizuno2017Continuum}.
Regarding their spatial structure, the QLVs consist of an unstable core and stable far-field components~\cite{Shimada2018Spatial}, and the latter decay algebraically in space~\cite{Lerner2016Statistics,Gartner2016Nonlinear}.
The vDOS of the QLVs follows a power law $g_{\mathrm{QLV}}(\omega)\sim\omega^\beta$, with $\beta=4$ having been observed in many glasses~\cite{Lerner2016Statistics, Mizuno2017Continuum, Lerner2017Effect,Shimada2018Anomalous,Kapteijns2018,Wang2019}, although $\beta\simeq3$ has also been observed for glasses prepared via rapid quenching from high-temperature liquids~\cite{Lerner2017Effect}.
The QLVs have also been shown to be highly anharmonic~\cite{Taraskin1999Anharmonicity, Xu2010Anharmonic} and to play important roles in thermal relaxation~\cite{Oligschleger1999, Widmer-Cooper2009,Lerner2018Characteristic} and mechanical relaxation under shear~\cite{Maloney2006, Tanguy2010, Manning2011}.

In parallel with these simulation efforts, various theories have been proposed to describe the anomalous vibrations of glasses.
First, an {elasticity theory with quenched disorder} has been developed~\cite{Schirmacher1998,Schirmacher2006Thermal, Schirmacher2007Acoustic, Wyart2010Scaling, Degiuli2014Effects, Degiuli2014Force}, in which glasses are regarded as elastic media with spatially fluctuating stiffness and the equation of motion is solved under mean-field-like approximations.
This {theory} predicts that the vDOS follows the non-Debye scaling $g(\omega) \sim \omega^2$ around the BP frequency when the system is close to the {\it elastic instability} originating from the fluctuating stiffness~\cite{Schirmacher2007Acoustic,Degiuli2014Force}.
When the system is marginally stable, the non-Debye scaling extends down to zero frequency.
Second, a microscopic replica theory has been developed~\cite{Franz2015Universal}.
This theory regards the perceptron model as a toy model that belongs to the same universality class as jammed systems do and studies the energy landscape of this model.
The theory reveals that this model has a marginally stable glass phase characterized by full replica symmetry breaking and that the vDOS follows the non-Debye scaling $g(\omega) \sim \omega^2$ in this phase.
Therefore, these two theories consistently predict a non-Debye scaling that extends down to zero frequency in marginally stable glasses.

However, these theoretical results are in sharp contrast to the simulation results, which show that in real glasses, the non-Debye scaling breaks down and QLVs appear in the low-frequency region.
One way to explain this observation is to suppose that real glasses are almost, but not exactly, marginally stable~\cite{Degiuli2014Effects, Lerner2014Breakdown}.
However, this approach does not explain the presence of the QLVs because the theories predict only (weakly perturbed) phonons in the corresponding frequency region~\cite{Schirmacher2006Thermal, Schirmacher2007Acoustic, Wyart2010Scaling, Degiuli2014Effects, Degiuli2014Force}.
Therefore, a new explanation for the low-frequency vibrations of glasses is necessary.
One recent attempt along these lines is to introduce fluctuations in the marginal stability~\cite{Ikeda2018Note, Ikeda2018Universal}.

In this Letter, we show that vibrations similar to QLVs can be directly derived from elasticity theory with quenched disorder.
Specifically, we find that the mechanical response to a local dipolar force becomes unstable in certain situations and gives rise to another contribution to the vDOS, one that has been overlooked thus far.
We show that the physical properties of these vibrations are fully consistent with those of QLVs.
{We also construct a simple elasticity model in which the additional contribution to the vDOS follows a quartic law. }

{\it Model.} ---To study the vibrational properties of glasses, we consider a simple spring network model consisting of a $d$-dimensional lattice of $N$ elements of unit mass~\cite{Schirmacher2007Acoustic,Wyart2010Scaling,Kohler2013Coherent,Degiuli2014Effects}.
The coordination number per element, $z_0$, is larger than $2d$.
Each nearest neighbor pair $\left<ij\right>$ has a spring whose stiffness $k_{ij}$ is an independent random variable obeying the probability distribution $P\left(k_{ij}\right)$.
The equation of motion is
\begin{equation}\label{eq:equation of motion}
  \frac{d^2}{dt^2}\boldsymbol{u}_i = -\sum_{\left<ij\right>}k_{ij}\left(\boldsymbol{u}_i-\boldsymbol{u}_j\right)\cdot\boldsymbol{n}_{ij},
\end{equation}
where $\boldsymbol{u}_i$ is the displacement of the $i$-th element and $\boldsymbol{n}_{ij}$ is the unit vector from the $j$-th element to the $i$-th element.
This equation can also be written in bra--ket notation as follows:
\begin{equation}
    \begin{split}
        \frac{d^2}{dt^2}\ket{\boldsymbol{u}} &= -\sum_{\left<ij\right>}k_{ij}\left(\ket{i}-\ket{j}\right)\boldsymbol{n}_{ij}\otimes\boldsymbol{n}_{ij}\left(\bra{i}-\bra{j}\right)\ket{\boldsymbol{u}} \\
        &\equiv - \hat{\mathcal{M}}\ket{\boldsymbol{u}},
    \end{split} 
\end{equation}
where $\braket{i|\boldsymbol{u}} = \boldsymbol{u}_i$ and $\hat{\mathcal{M}}$ is the dynamical matrix.

{\it Effective medium approximation}. ---We employ the effective medium approximation (EMA)~\cite{Schirmacher1998,Wyart2010Scaling, Kohler2013Coherent, Degiuli2014Effects} to analyze the vibrational properties of the present model.
This approximation yields the approximate disorder-averaged Green function $\overline{\hat{\mathcal{G}}(\omega)} = \overline{(\hat{\mathcal{M}}-\omega^2)^{-1}}$ within a mean-field-like approach.
First, we decompose the dynamical matrix as follows:
\begin{equation}\label{decomposition}
  \begin{split}
    \hat{\mathcal{M}} - \omega^2 &= \sum_{\alpha=\langle i j\rangle}k_{\alpha}\ket{\alpha}\boldsymbol{n}_\alpha\otimes\boldsymbol{n}_\alpha\bra{\alpha} - \omega^2 \\
    &= \left[k_{\mathrm{eff}}(\omega)\sum_{\alpha=\langle i j\rangle}\ket{\alpha}\boldsymbol{n}_\alpha\otimes\boldsymbol{n}_\alpha\bra{\alpha} - \omega^2\right] \\
    &+ \sum_{\alpha=\langle i j\rangle}\left[k_{\alpha}-k_{\mathrm{eff}}(\omega)\right]\ket{\alpha}\boldsymbol{n}_\alpha\otimes\boldsymbol{n}_\alpha\bra{\alpha} \\
    &\equiv \hat{\mathcal{G}}_{\mathrm{eff}}(\omega)^{-1} + \hat{\mathcal{V}}(\omega),
  \end{split}
\end{equation}
where $\ket{\alpha}=\ket{i}-\ket{j}$.
Since the effective stiffness $k_{\mathrm{eff}}(\omega)$ is independent of space, $\hat{\mathcal{G}}_{\mathrm{eff}}(\omega)$ is simply the Green function for the homogeneous system.
Instead of using the exact form for $\hat{\mathcal{G}}_{\mathrm{eff}}(\omega)$, we adopt the same approximate Green function used in Ref.~\cite{Wyart2010Scaling, Degiuli2014Effects}:
\begin{equation}\label{eq:unperturbed Green function}
  \begin{split}
    \bra{i}\hat{\mathcal{G}}_{\mathrm{eff}}(\omega)\ket{j} &\equiv \hat{G}_{\mathrm{eff}}(\boldsymbol{r}_{ij},\omega) \\
    &= \int_{0<|\boldsymbol{q}|<q_D}\frac{d\boldsymbol{q}}{(2\pi)^{d}}\frac{e^{i\boldsymbol{q}\cdot\boldsymbol{r}_{ij}}}{k_{\mathrm{eff}}(\omega)\boldsymbol{q}^{2}-\omega^{2}}\hat{\delta}_d,
  \end{split}
\end{equation}
where $\boldsymbol{r}_{ij}$ is the vector from the $j$-th element to the $i$-th element, $\hat{\delta}_d$ is the $d\times d$ identity matrix, and $q_D$ is the Debye wavenumber determined by $1 = \int_{0<|\boldsymbol{q}|<q_D}\frac{d\boldsymbol{q}}{\left(2\pi\right)^d}$.
{Note that we set the number density to $\rho=1$.}

Treating the second term $\hat{\mathcal{V}}(\omega)$ as a perturbation, we can write the transfer matrix as
\begin{equation}\label{eq:T matrix series}
  \begin{split}
    \hat{\mathcal{T}}(\omega) &= -\hat{\mathcal{V}}(\omega)\left[1 + \hat{\mathcal{G}}_{\mathrm{eff}}(\omega)\hat{\mathcal{V}}(\omega)\right]^{-1} \\
    &= \sum_{\alpha=\left<ij\right>} \hat{\mathcal{T}}_{\alpha}(\omega) + \sum_{\alpha=\left<ij\right>} \sum_{\beta\neq\alpha} \hat{\mathcal{T}}_{\alpha}(\omega) \hat{\mathcal{G}}_{\mathrm{eff}}(\omega)\hat{\mathcal{T}}_{\beta}(\omega) + \cdots,
  \end{split}
\end{equation}
where
\begin{equation}\label{eq:T matrix}
  \hat{\mathcal{T}}_{\alpha}(\omega) = \frac{\left[k_{\mathrm{eff}}(\omega)- k_{\alpha}\right]\ket{\alpha}\boldsymbol{n}_\alpha\otimes\boldsymbol{n}_\alpha\bra{\alpha}}{1 - \left[k_{\mathrm{eff}}(\omega)- k_{\alpha}\right] \boldsymbol{n}_\alpha^T\bra{\alpha}\hat{\mathcal{G}}_{\mathrm{eff}}(\omega)\ket{\alpha}\boldsymbol{n}_\alpha}.
\end{equation}
The EMA self-consistent equation for the effective stiffness is derived from the condition $\overline{\hat{\mathcal{T}}_\alpha(\omega)} = 0$.
This finally leads to the equation
{
\begin{equation}\label{eq:finite frequency self consistent equation}
    \overline{\frac{k_{\mathrm{eff}}(\omega)- k_{\alpha}}{k_{\mathrm{eff}}(\omega) - \left[k_{\mathrm{eff}}(\omega)- k_{\alpha}\right]\theta\left[1 + \omega^2G(\omega)\right]}} = 0,
\end{equation}
}
\noindent where $\theta=2d/z_0<1$ and we have used the identity derived from homogeneity and isotropy~\cite{Wyart2010Scaling, Degiuli2014Effects}.
$G(\omega)$ is defined as
\begin{equation}\label{eq:Green function}
  G(\omega) = \int_{0<|\boldsymbol{q}|<q_D}\frac{d\boldsymbol{q}}{(2\pi)^{d}}\frac{e^{i\boldsymbol{q}\cdot\boldsymbol{r}}}{k_{\mathrm{eff}}(\omega)\boldsymbol{q}^{2}-\omega^{2}}.
\end{equation}
The vibrational properties of the model can now be calculated by solving Eq.~(\ref{eq:finite frequency self consistent equation}) using $P(k_{ij})$ and $\theta$ as the sole inputs.
Generally, the effective stiffness is a complex number, $k_{\mathrm{eff}}(\omega) = k_r(\omega)-i\Sigma(\omega)$.
The elastic stability of the system can be seen from the imaginary part at zero frequency: $\Sigma(0)>0$ when the system is unstable.
The imaginary part of $G(\omega)$ yields the vDOS
\begin{equation}
  g(\omega) = \frac{2d\omega}{\pi}\Im G(\omega).
\end{equation}

{\it Conventional instability.} ---The EMA self-consistent equation has been solved for several specific $P(k_{ij})$, and the emergence of elastic instability under various situations has been reported~\cite{Schirmacher2007Acoustic,Kohler2013Coherent,Degiuli2014Effects, Degiuli2014Force}.
Remarkably, the vDOS has been shown to follow the non-Debye scaling $g(\omega) \sim \omega^2$ near this instability~\cite{Schirmacher2007Acoustic,Degiuli2014Effects}.
Here, we will first show that the emergence of non-Debye scaling is a robust property of the present model when $\theta \ll 1$.
We will present a brief sketch of the calculation (see the Supplemental Material for more details).
When $\theta\ll1$, Eq.~(\ref{eq:finite frequency self consistent equation}) reduces to
\begin{equation}\label{eq:self consistent equation for uniform distribution}
  k_{\mathrm{eff}}(\omega)^3 -\mu k_{\mathrm{eff}}(\omega)^2 + \sigma^2\theta k_{\mathrm{eff}}(\omega) + \sigma^2\theta A_d\omega^2 = 0, 
\end{equation}
where $\mu$ and $\sigma^2$ are the mean and variance, respectively, of the distribution: $\mu \equiv \int dk_{ij}k_{ij}P\left(k_{ij}\right)$ and $\sigma^2 \equiv \int dk_{ij}k_{ij}^2P\left(k_{ij}\right) - \mu^2$.
At $\omega=0$, this equation further simplifies to a quadratic equation, and $\sigma_c\equiv\mu/2\sqrt{\theta}$ is the critical value above which the system is unstable, $\Sigma(0)>0$.
When $\omega\neq0$, we can approximately solve Eq.~(\ref{eq:self consistent equation for uniform distribution}) under the condition $(\sigma_c^2-\sigma^2)\theta/\mu^2 \sim \omega^2/\mu \ll 1$, and the resulting solution is
\begin{equation}\label{eq:solution of the conventional instability}
  k_{\mathrm{eff}}(\omega) = \frac{\mu}{2} - i\sqrt{\frac{\mu}{2}}\sqrt{A_d\omega^2 - A_d{\omega_0}^2},
\end{equation}
where
\begin{equation}
  \omega_0 \equiv \sqrt{\frac{2\theta}{\mu A_d}}\sqrt{\sigma_c^2-\sigma^2}.
\end{equation}
The vDOS is written as
\begin{equation}
    g(\omega) \sim \omega\sqrt{A_d\omega^2 - A_d{\omega_0}^2} \sim \omega^2.
\end{equation}
This result is valid for all probability distributions $P(k_{ij})$ {as long as the moments are finite}.

The instability condition $\sigma = \sigma_c \sim \mu/\sqrt{\theta}$ has a simple physical interpretation.
The mean and standard deviation of the sum of the stiffnesses of the springs connected to an element per degree of freedom of that element are on the orders of $\mu/\theta$ and $\sigma/\sqrt{\theta}$, respectively.
When they are of the same order, $\sigma \sim \mu/\sqrt{\theta}$, the system becomes unstable.
We call this instability the {\it conventional instability}.
The above derivation suggests that the conventional instability appears when the self-consistent equation reduces to an algebraic equation.
In our case, the condition $\theta\ll1$ plays a key role, and it corresponds to the large-dimension limit of the scalar displacement version of the model, which is often used to describe the low-frequency vibrations of glasses~\cite{Schirmacher2007Acoustic,Kohler2013Coherent}.
We discuss this point in detail in Ref.~\cite{Shimada_inprep}.

{\it Another instability induced by the local response.} ---When $\theta$ is not small, the derivation above does not apply.
We will now show that another type of instability emerges in such a situation.
We focus on the zero-frequency case to discuss the elastic stability.
At $\omega=0$, Eq.~(\ref{eq:finite frequency self consistent equation}) is
\begin{equation}\label{eq:zero frequency self consistent equation}
  \int\frac{dk_\alpha P(k_\alpha)}{k_\alpha + (1-\theta)k_{\mathrm{eff}}(0)/\theta} = \frac{1}{k_{\mathrm{eff}}(0)}.
\end{equation}
Suppose that the system is stable; the effective stiffness consists of the positive real part $k_{\mathrm{eff}}(0)$ and an infinitesimally small imaginary part $-i\epsilon$.
Then, the left-hand side of Eq.~(\ref{eq:zero frequency self consistent equation}) becomes
\begin{equation}
  \mathcal{P}\int\frac{dk_\alpha P(k_\alpha)}{k_\alpha + (1-\theta)k_{\mathrm{eff}}(0)/\theta} + i\pi P\left[-\left(1-\theta\right)k_{\mathrm{eff}}(0)/\theta\right],
\end{equation}
where $\mathcal{P}$ denotes the Cauchy principal value.
By inserting this expression into Eq.~(\ref{eq:zero frequency self consistent equation}) and focusing on the imaginary part, one finds that
\begin{equation}\label{eq:stability condition}
  P\left[-\left(1-\theta\right)k_{\mathrm{eff}}(0)/\theta\right] = 0
\end{equation}
is necessary for the effective stiffness to be real.
Thus, violation of this condition causes instability.
This condition is generally different from those for the conventional instability discussed in the previous section.

This instability has a simple physical interpretation.
To illustrate this, we consider a ``defect'' model in which all the springs have the same real stiffness $k_{\mathrm{eff}}(0)>0$ except for one defect spring, whose stiffness is $k_\alpha$.
The dynamical matrix of this defect model is
\begin{equation}\label{M of defect model}
    \begin{split}
        \hat{\mathcal{M}} &= k_{\mathrm{eff}}(0)\sum_{\beta=\langle k l\rangle}\ket{\beta}\boldsymbol{n}_\beta\otimes\boldsymbol{n}_\beta\bra{\beta} \\
        &+ \left[k_{\alpha}-k_{\mathrm{eff}}(0)\right]\ket{\alpha}\boldsymbol{n}_\alpha\otimes\boldsymbol{n}_\alpha\bra{\alpha}.
    \end{split}
\end{equation}
By calculating the transfer matrix for this dynamical matrix, one can see that only the first term of the expansion in Eq.~(\ref{eq:T matrix series}) remains and that the total Green function is exactly
\begin{equation}
    \hat{\mathcal{G}}(\omega) = \hat{\mathcal{G}}_0(\omega) + \hat{\mathcal{G}}_0(\omega)\hat{\mathcal{T}}_{\alpha0}(\omega)\hat{\mathcal{G}}_0(\omega),
\end{equation}
where $\hat{\mathcal{G}}_0(\omega)$ is the unperturbed Green function with stiffness $k_{\mathrm{eff}}(0)$ and $\hat{\mathcal{T}}_{\alpha0}(\omega)$ is Eq.~(\ref{eq:T matrix}) with $k_{\mathrm{eff}}(\omega)\to k_{\mathrm{eff}}(0)$ and $\hat{\mathcal{G}}_{\mathrm{eff}}(\omega)\to\hat{\mathcal{G}}_{0}(\omega)$.
As seen from Eq.~(\ref{eq:T matrix}), $\hat{\mathcal{T}}_{\alpha0}(\omega)$ diverges when $k_\alpha=-(1-\theta)k_{\mathrm{eff}}(0)/\theta$, which means that the system has a nontrivial zero mode.
This zero mode can be constructed (without normalization) as follows:
\begin{equation}
    \ket{0} \equiv \hat{\mathcal{G}}_0(0)\ket{\alpha}\boldsymbol{n}_\alpha.
\end{equation}
The eigenrelation $\hat{\mathcal{M}}\ket{0} = 0$ can be checked as follows.
Taking the product of the first term of Eq.~(\ref{M of defect model}) by $\ket{0}$ yields $\ket{\alpha}\boldsymbol{n}_\alpha$.
The second term yields
\begin{equation}
    \begin{split}
        &\left[k_{\alpha}-k_{\mathrm{eff}}(0)\right]\ket{\alpha}\boldsymbol{n}_\alpha\otimes\boldsymbol{n}_\alpha\braket{\alpha|0} \\
        &= \left[k_{\alpha}-k_{\mathrm{eff}}(0)\right]\ket{\alpha}\boldsymbol{n}_\alpha\boldsymbol{n}_\alpha^T\bra{\alpha}\hat{\mathcal{G}}_0(0)\ket{\alpha}\boldsymbol{n}_\alpha \\
        &= \frac{\theta}{k_{\mathrm{eff}}}\left[k_{\alpha}-k_{\mathrm{eff}}(0)\right]\ket{\alpha}\boldsymbol{n}_\alpha, 
    \end{split}
\end{equation}
where we have used the identity derived from homogeneity and isotropy~\cite{Wyart2010Scaling, Degiuli2014Effects}.
Then, we obtain
\begin{equation}\label{M0}
    \hat{\mathcal{M}}\ket{0} = \left\{ 1 + \frac{\theta [k_\alpha - k_{\mathrm{eff}}(0)]}{k_{\mathrm{eff}}(0)} \right\} \ket{\alpha}\boldsymbol{n}_\alpha.
\end{equation}
Therefore, $\hat{\mathcal{M}}\ket{0} = 0$ when the ``defect'' bond $\alpha$ has a negative stiffness $k_\alpha = -(1-\theta)k_{\mathrm{eff}}(0)/\theta$.
Furthermore, when $k_\alpha < -(1-\theta)k_{\mathrm{eff}}(0)/\theta$, the system becomes unstable along the $\ket{0}$ direction.
Thus, the stability condition in Eq.~(\ref{eq:stability condition}) can be interpreted as follows:
when Eq.~(\ref{eq:stability condition}) is violated, an extensive number of springs become ``defects'' and unstable modes appear.
The instability identified here is completely different from the conventional instability discussed in the previous section.
Since this instability is local in nature, we call it the {\it local instability}.
Note that models in which $P(k_{ij})$ is supported on the whole real axis, such as the Gaussian distribution, are always unstable in terms of this local instability.

The mode associated with the local instability, $\ket{0}$, is the response to a local dipolar force in the unperturbed system.
The relationship between the QLVs and the response to a local dipolar force in glasses has recently been revealed~\cite{Lerner2014Breakdown,Yan2016On, Lerner2018Characteristic, Shimada2018Spatial}.
First, the vibrational amplitude $|\braket{i|\lambda}|^2$ for the QLVs $\ket{\lambda}$ has been shown to have an asymptotic decay profile {$r^{2(1-d)}$}, which is exactly the same as the response to a local dipolar force $|\braket{i|0}|^2$~\cite{Lerner2016Statistics}.
Second, the spatial extent of the core part of the QLVs is comparable to that of the response to a local dipolar force~\cite{Yan2016On,Shimada2018Spatial}.
Furthermore, the energetics of the QLVs are similar to those of $\ket{0}$: the QLVs consist of an energetically unstable core and far-field components that stabilize the modes~\cite{Shimada2018Spatial}, and $\ket{0}$ also has a local unstable part (the negative second term in Eq.~(\ref{M0})) that is stabilized by a far-field component (the positive first term in Eq.~(\ref{M0})).
Therefore, we conclude that the mode associated with the local instability exhibits strong similarities with the QLVs.

\begin{figure}
    \centering
    \includegraphics[width=0.45\textwidth]{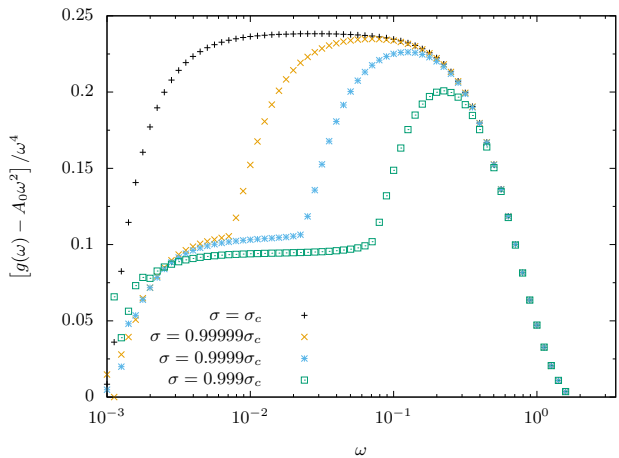}
\caption{
The deviation of the vDOS $g(\omega)$ from the Debye law $g_{\mathrm{Debye}}(\omega) = A_0\omega^2$.
To emphasize the frequency dependence, we divide by $\omega^4$.
$\sigma_c$ is a numerically estimated critical value, up to which Eq.~(\ref{eq:stability condition}) holds.
Although the data for $\sigma=\sigma_c$ deviate from the plateau in the lowest-frequency region, this deviation is due to numerical errors in estimating $\sigma_c$.
}
    \label{fig:fig1}
\end{figure}

{\it Toy model for the quartic law.} ---Up to this point, we have seen that the response to a local dipolar force leads to local instability.
The next question is whether this local instability reproduces the vDOS of the QLVs, $g_{\mathrm{QLV}}(\omega)\sim\omega^4$.
To answer this question, we consider the following distribution:
\begin{equation}~\label{eq:distribution for the QLVs}
    P\left(k_{\alpha}\right)=\left\{\begin{array}{ll}{C\left(k_{\alpha}-\mu+\Delta\right)^{3 / 2}} & {\left(\mu-\Delta<k_{\alpha}<\mu\right)} \\ {C\left(-k_{\alpha}+\mu+\Delta\right)^{3 / 2}} & {\left(\mu<k_{\alpha}<\mu+\Delta\right)} \\ 0 & (\mathrm{otherwise})\end{array}\right. ,
\end{equation}
where $C=5\Delta^{-5/2}/4$ and the variance is $\sigma^2={8 \Delta^{2}}/{63}$.
The key feature of this distribution is the power law in the tail, and the detailed shape does not affect the following observations~\cite{Shimada_inprep}.

We numerically solve Eq.~(\ref{eq:finite frequency self consistent equation}) with Eq.~(\ref{eq:distribution for the QLVs}) for $d=3$ and find that the local instability arises at {$\sigma=\sigma_c\sim 0.356$}, i.e., Eq.~(\ref{eq:stability condition}) holds only up to $\sigma=\sigma_c$, above which the system is unstable, $\Sigma(0)>0$.
To solve the equation numerically, we transform it into a simpler form, as detailed in the Supplemental Material.
The low-frequency part of the full vDOS $g(\omega)$ consists of a Debye law part, $g_{\mathrm{Debye}}(\omega)=A_0\omega^{d-1}=A_0\omega^2$, and a deviation due to disorder.
To highlight the deviation, we subtract $g_{\mathrm{Debye}}(\omega)$ from $g(\omega)$ and then divide by $\omega^4$.
We plot the results for $\sigma=\sigma_c$, $0.99999\sigma_c$, $0.9999\sigma_c$, and $0.999\sigma_c$ with $\theta=0.9999$ in Fig.~\ref{fig:fig1}.
Two plateaus appear in the low-frequency region.
The appearance of a plateau means that $g(\omega) = g_{\mathrm{Debye}}(\omega) + \alpha \omega^4$, where the plateau height corresponds to the prefactor $\alpha$.
When $\sigma$ is far from $\sigma_c$, the lower plateau (smaller $\alpha$) dominates the low-frequency region.
This $\omega^4$ law is due to Rayleigh scattering, $g(\omega)\sim\omega^{d+1}$, and has been observed in all elasticity models with perturbations~\cite{John1983Localization, Schirmacher2006Thermal, Kohler2013Coherent, Wyart2010Scaling, Degiuli2014Effects, Degiuli2014Force}.
This phenomenon cannot be identified as the QLVs since the QLVs follow the $\omega^4$ law even for $d \neq 3$~\cite{Kapteijns2018Universal}.
On the other hand, when $\sigma$ is approaching $\sigma_c$, a higher plateau (larger $\alpha$) appears.
These modes are caused by the proximity of the local instability.
Notably, the higher plateau extends down to zero frequency as $\sigma \to \sigma_c$, which means that the local instability induces gapless vibrational modes following the $\omega^4$ law, as observed for the QLVs.
Note that the deviation of the data from the plateau for $\sigma=\sigma_c$ in the lowest-frequency region is simply due to numerical errors in estimating $\sigma_c$.

Since the vDOS is related to the scattering of phonons as described by the imaginary part of the effective stiffness, $\Sigma(\omega)=-\Im k_{\mathrm{eff}}(\omega)$~\cite{Schirmacher2006Thermal,Schirmacher2007Acoustic,Wyart2010Scaling,Degiuli2014Effects,Degiuli2014Force,Mizuno2018Phonon}, the prefactor $\alpha$ is directly related to the phonon scattering amplitude.
Thus, the increase in $\alpha$ due to the local instability implies the enhancement of phonon scattering, which is consistent with the numerical observation of stronger scattering due to the QLVs~\cite{Mizuno2018Phonon}.

{\it Summary and Discussion.} ---We have found a new instability in elasticity theory with quenched disorder.
This instability corresponds to the instability of the response to a local dipolar force, and hence, we call it the local instability.
We have shown that the spatial structure and energetics of the mode associated with this local instability are fully consistent with those of the QLVs of glasses.
By analyzing a toy model, we have shown that the local instability induces an additional contribution to the vDOS, which can reproduce the $\omega^4$ law characteristic of the QLVs.
These results offer a new way to understand and predict the low-frequency vibrational properties of glasses.

Our results can be used to refine the concept of the marginal stability of glasses.
Since the non-Debye scaling breaks down at a finite frequency, real glasses are considered to be almost, but not exactly, marginally stable~\cite{Degiuli2014Effects,Lerner2014Breakdown}.
However, our findings suggest that glasses potentially have two different instabilities: the conventional instability and the local instability.
If the system is at a finite distance from the local instability, the prefactor of the quartic frequency dependence in the vDOS decreases in magnitude, as shown in Fig.~\ref{fig:fig1}.
No such drop has ever been observed in numerical simulations, however, and QLVs exist down to zero frequency~\cite{Lerner2016Statistics, Mizuno2017Continuum, Shimada2018Anomalous}.
This fact indicates that real glasses are marginally stable in terms of the local instability, not the conventional instability.
This is also consistent with observations of real glasses under shear, in which the onset strain for small plastic deformations decreases without bound with increasing system size~\cite{Karmakar2010a}.

In the present work, we have limited our analysis to the EMA.
When one goes beyond the EMA, various kinds of instabilities should be allowed due to different combinations of $\hat{\mathcal{T}}_\alpha(\omega)$.
However, the vibrational modes associated with such instabilities have not been detected in either numerical or experimental studies of glasses.
The only low-frequency modes of glasses detected so far are the QLVs, whose spatial structure is similar to that of the response to a local dipolar force.
Therefore, we expect that the random spring network under the EMA should be a suitable framework for the analysis and that the inclusion of higher-order terms will not improve the description of the vibrational properties of glasses~\footnote{
Note that in Refs.~\cite{Degiuli2014Effects,Degiuli2014Force}, the initial stress was considered to be the source of the instability.
Specifically, we need the term $-\frac{f_\alpha}{|\boldsymbol{r}_\alpha|}\ket{\alpha}\left(\hat{\delta}_d- \boldsymbol{n}_\alpha\otimes\boldsymbol{n}_\alpha\right)\bra{\alpha}$, where $f_\alpha$ is a force between a pair $\alpha$, in the dynamical matrix.
Although we have neglected the initial stress in this paper, the mechanism of the local instability is general and valid even if the initial stress is considered.
In this case, the distribution of $f_\alpha$ plays a central role.
}.

In our forthcoming paper~\cite{Shimada_inprep}, we will discuss further details of the model.
We will prove that when the tail of the distribution $P(k_\alpha)$ follows a power law with an exponent $\nu$, this leads to a power law in the vDOS, $g(\omega)\sim\omega^{2\nu+1}$.
The last result of this paper is an example of this general relation.
{Furthermore, we will discuss the fact that our model can be interpreted as a coarse-grained model and the fact that its coarse-graining length scale is of the same order as the length of the QLVs.}

{\it Acknowledgments.} ---We thank C.~Dasgupta, C.~Hotta, K.~Hukushima, H.~Ikeda, S.~Ishihara, S.~Yaida, and F.~Zamponi for fruitful discussions.
This work was supported by JSPS KAKENHI Grant Numbers 19J20036, 17K14369, 17H04853, 16H04034, 18H05225, 19K14670, and 19H01812.
This work was also partially supported by the Asahi Glass Foundation.

\bibliographystyle{unsrt}
\bibliography{manuscript}

\clearpage

\onecolumngrid
\appendix

\section*{Supplemental material}

In this supplemental material, we explain the details of our calculations for deriving (i) the solution for the quadratic law governing the conventional instability and (ii) the simpler form of the self-consistent equation for the quartic law governing the local instability.

\subsection*{Solution for the quadratic law}

We solve Eq.~(\ref{eq:finite frequency self consistent equation}) with a distribution with finite moments under the condition $\theta\ll1$.
In particular, the mean is $\mu > 0$, and the variance is $\sigma^2$.
Equation~(\ref{eq:finite frequency self consistent equation}) can be expanded as follows:
\begin{equation}
    k_{\mathrm{eff}}(\omega) - \overline{k_\alpha} + \overline{\left[k_{\mathrm{eff}}(\omega)-k_\alpha\right]^2}\frac{\theta}{k_{\mathrm{eff}}(\omega)}\left[1 + \omega^2G(\omega)\right] \simeq 0.
\end{equation}
By averaging over $k_\alpha$, we obtain
\begin{equation}\label{eq:self consistent equation for the general distribution}
  \begin{split}
    k_{\mathrm{eff}}(\omega)^2 - \mu k_{\mathrm{eff}}(\omega) + \theta\left[k_{\mathrm{eff}}(\omega)^2 - 2\mu k_{\mathrm{eff}}(\omega) + (\sigma^2+\mu^2)\right]\left[1 + \omega^2G(\omega)\right] &= 0, \\
    k_{\mathrm{eff}}(\omega)^2 - \mu k_{\mathrm{eff}}(\omega) + \theta(\sigma^2+\mu^2) + \theta\omega^2G(\omega)\left[k_{\mathrm{eff}}(\omega)^2 - 2\mu k_{\mathrm{eff}}(\omega) + (\sigma^2+\mu^2)\right] &\simeq 0,\\
  \end{split}
\end{equation}
where in the last line, we neglect small terms.
At zero frequency, the equation further reduces to a quadratic equation, and its solution has a critical value at $\sigma_c\equiv\mu/2\sqrt{\theta}\gg\mu$, above which the system is unstable, $\Sigma(0)>0$.

Since we are interested in vibrations near $\sigma=\sigma_c$, we assume that $\sigma\gg\mu$ and derive an approximate expression for $k_{\mathrm{eff}}(\omega)$.
We focus only on the lowest-frequency region and approximate $G(\omega)$ as
\begin{equation}\label{eq:approximate Green function}
  \begin{split}
    G(\omega) &\simeq \frac{1}{k_{\mathrm{eff}}(\omega)}\int\frac{d\boldsymbol{q}}{(2\pi)^d}\frac{1}{\boldsymbol{q}^2} \\
    &= \frac{S_{d-1}}{k_{\mathrm{eff}}(\omega)(2\pi)^d}\int_0^{q_D}dq q^{d-3} \\
    &= \frac{1}{k_{\mathrm{eff}}(\omega)}\frac{d}{(d-2)q_D^2} \equiv \frac{A_d}{k_{\mathrm{eff}}(\omega)},
  \end{split}
\end{equation}
where from the second line to the third line, we use the definition of $q_D$.
Note that we neglect the contribution from the pole, which is responsible for the Debye law and Rayleigh scattering~\cite{Degiuli2014Force}.
By substituting Eq.~(\ref{eq:approximate Green function}) into Eq.~(\ref{eq:self consistent equation for the general distribution}), we obtain Eq.~(\ref{eq:self consistent equation for uniform distribution}), as follows:
\begin{equation}
    \begin{split}
        k_{\mathrm{eff}}(\omega)^2 - \mu k_{\mathrm{eff}}(\omega) + \theta(\sigma^2+\mu^2) + \theta\omega^2\frac{A_d}{k_{\mathrm{eff}}(\omega)}\left[k_{\mathrm{eff}}(\omega)^2 - 2\mu k_{\mathrm{eff}}(\omega) + (\sigma^2+\mu^2)\right] &= 0, \\
        k_{\mathrm{eff}}(\omega)^3 - \mu k_{\mathrm{eff}}(\omega)^2 + \theta\sigma^2k_{\mathrm{eff}}(\omega) + \theta A_d \omega^2 \sigma^2 &\simeq 0, \\
    \end{split} 
\end{equation}
where we use the condition $\sigma\gg\mu$.

To solve Eq.~(\ref{eq:self consistent equation for uniform distribution}), we substitute $k_{\mathrm{eff}}(\omega) = y + \mu/3$ into the above expression:
\begin{equation}
  \begin{split}
    & k_{\mathrm{eff}}(\omega)^3 -\mu k_{\mathrm{eff}}(\omega)^2 + \sigma^2\theta k_{\mathrm{eff}}(\omega) + \sigma^2\theta A_d\omega^2 \\
    &= y^3 + 3\left(-\frac{\mu^2}{9} + \frac{\sigma^2\theta}{3} \right)y  + 2\left(- \frac{\mu^3}{27} + \mu\frac{\sigma^2\theta}{6}  + \frac{\sigma^2\theta}{2} A_d\omega^2\right) \\
    &\equiv y^3 + 3Py + 2Q.
  \end{split}
\end{equation}
Using the critical value defined above, $P$ can be written as
\begin{equation}
    \begin{split}
    P 
    &= -\left(\frac{\mu}{6}\right)^2 - \frac{\sigma_c^2-\sigma^2}{3}\theta \\
    &\equiv -\left(\frac{\mu}{6}\right)^2 - \frac{\delta_\sigma^2\theta}{3}. \\
  \end{split}
\end{equation}
$Q$ becomes
\begin{equation}
  \begin{split}
    Q 
    &= \left(\frac{\mu}{6}\right)^3 - \mu\frac{\sigma_c^2 - \sigma^2}{6}\theta + \frac{\sigma^2\theta}{2} A_d\omega^2  \\
    &\equiv \left(\frac{\mu}{6}\right)^3 - \frac{\mu\delta_\sigma^2}{6}\theta + \frac{\sigma^2\theta}{2}A_d\omega^2.  \\
  \end{split} 
\end{equation}
For our purposes, it suffices to assume that $\theta\delta_\sigma^2/\mu^2\sim\omega^2/\mu \ll 1$ and to regard these terms as perturbations.
Thus, $Q$ can be approximated as
\begin{equation}
  Q \simeq \left(\frac{\mu}{6}\right)^3 - \frac{\mu\delta_\sigma^2}{6}\theta + \frac{\sigma_c^2\theta}{2}A_d\omega^2.
\end{equation}
Next, we need to compute $\left(-Q \pm \sqrt{Q^2 + P^3}\right)^{1/3}$.
Let us first consider the terms under the square root:
\begin{equation}
  \begin{split}
     Q^2 + P^3 
    &=  - 3\left(\frac{\mu}{6}\right)^4{\delta_\sigma^2}{\theta} + \frac{2}{3}\left(\frac{\mu\delta_\sigma^2}{6}\theta\right)^2 + \frac{1}{4}\left(\sigma_c^2\theta A_d\omega^2\right)^2 + \left(\frac{\mu}{6}\right)^3\sigma_c^2\theta A_d\omega^2 - \frac{\mu\delta_\sigma^2}{6}\sigma_c^2\theta^2A_d\omega^2 \\
    &\simeq  - 3\left(\frac{\mu}{6}\right)^4{\delta_\sigma^2}{\theta} + \left(\frac{\mu}{6}\right)^3{\sigma_c^2}{\theta}A_d\omega^2  \\
    &= 9\left(\frac{\mu}{6}\right)^5\left(A_d\omega^2　- 2\frac{\delta_\sigma^2\theta}{\mu}\right).  \\
  \end{split}
\end{equation}
As a result,
\begin{equation}
  \begin{split}
    \left(-Q \pm \sqrt{Q^2 + P^3}\right)^{1/3} &= \left[\left(-\frac{\mu}{6}\right)^3 + \frac{\mu\delta_\sigma^2}{6}\theta - \frac{\sigma_c^2}{2}\theta A_d\omega^2 \pm \sqrt{9\left(\frac{\mu}{6}\right)^5\left(A_d\omega^2　- 2\frac{\delta_\sigma^2\theta}{\mu}\right)}\right]^{1/3} \\
    &\simeq -\frac{\mu}{6}\left[1 \pm \frac{-6}{\mu}\sqrt{\frac{\mu}{6}\left(A_d\omega^2　- 2\frac{\delta_\sigma^2\theta}{\mu}\right)}\right] \\
    &= -\frac{\mu}{6} \pm \sqrt{\frac{\mu}{6}\left(A_d\omega^2　- 2\frac{\delta_\sigma^2\theta}{\mu}\right)}.
  \end{split}
\end{equation}
We choose the solution that satisfies $k_{\mathrm{eff}}(0)\to\mu$ as $\sigma\to0$ and $\Sigma(\omega) \equiv -\Im k_{\mathrm{eff}}(\omega) > 0$.
Thus, we obtain
\begin{equation}
  \begin{split}
    k_{\mathrm{eff}}(\omega)  &= \frac{\mu}{3} + \frac{-1-\sqrt{3}i}{2}\left(-Q+\sqrt{Q^2 + P^3}\right)^{1/3} + \frac{-1+\sqrt{3}i}{2}\left(-Q-\sqrt{Q^2 + P^3}\right)^{1/3} \\
    &\simeq \frac{\mu}{3} + \frac{-1-\sqrt{3}i}{2}\left[-\frac{\mu}{6} + \sqrt{\frac{\mu}{6}\left(A_d\omega^2　- 2\frac{\delta_\sigma^2\theta}{\mu}\right)}\right] + \frac{-1+\sqrt{3}i}{2}\left[ -\frac{\mu}{6} - \sqrt{\frac{\mu}{6}\left(A_d\omega^2　- 2\frac{\delta_\sigma^2\theta}{\mu}\right)}\right] \\
    &= \frac{\mu}{2} - i\sqrt{\frac{\mu}{2}\left(A_d\omega^2　- 2\frac{\delta_\sigma^2\theta}{\mu}\right)} \equiv \frac{\mu}{2} - i\sqrt{\frac{\mu}{2}}\sqrt{A_d\omega^2　- A_d{\omega_0}^2}.
  \end{split}
\end{equation}
This is Eq.~(\ref{eq:solution of the conventional instability}).

\subsection*{Self-consistent equation for the quartic law}\label{sec:Self consistent equation for the QLVs}

Here, we simplify Eq.~(\ref{eq:finite frequency self consistent equation}) with the distribution given in Eq.~(\ref{eq:distribution for the QLVs}).
First, Eq.~(\ref{eq:finite frequency self consistent equation}) can be transformed into
\begin{equation}\label{eq:with kappa}
  \overline{ \frac{1}{k_\alpha + \kappa(\omega)} } = \theta\frac{1 + \omega^2G\left(\omega\right)}{k_{\mathrm{eff}}(\omega)}, 
\end{equation}
where
\begin{equation}
  \kappa(\omega) = \frac { \theta^{-1} -  1 - \omega ^ { 2 } { G } ( \omega )} { 1 + \omega ^ { 2 } { G } ( \omega ) } k _ { \mathrm{eff} }(\omega).
\end{equation}
The left-hand side of Eq.~(\ref{eq:with kappa}) becomes
\begin{equation}
    \begin{split}
        &\overline{\frac{1}{k_\alpha + \kappa(\omega)}} \\
        &= \frac{5}{4\Delta}\left[\int_{-1}^0dx\frac{(x+1)^{3/2}}{x + \left[\mu + \kappa(\omega)\right]/\Delta} + \int^1_0 dx\frac{(-x+1)^{3/2}}{x + \left[\mu + \kappa(\omega)\right]/\Delta}\right] \\
        &\equiv \frac{5}{4\Delta}\left[\int_{-1}^0dx\frac{(x+1)^{3/2}}{x + z} + \int^1_0 dx\frac{(-x+1)^{3/2}}{x + z}\right] \\
        &= \frac{5}{4\Delta}\left[\int_{-1}^0dx\frac{(x+z-z+1)\sqrt{x+1}}{x + z} + \int^1_0 dx\frac{(-x-z+z+1)\sqrt{-x+1}}{x + z} \right]\\
        &= \frac{5}{4\Delta}\left[-4z + 2(z-1)^{3/2}\int_0^{1/\sqrt{z-1}}\frac{dt}{t^2 + 1} - 2(z+1)^{3/2}\int_0^{1/\sqrt{z+1}}\frac{dt}{t^2-1}\right] \\
        &= \frac{5}{4\Delta}\left[-4z + 2(z-1)^{3/2}\arctan{\frac{1}{\sqrt{z-1}}} + 2(z+1)^{3/2}\arctanh{\frac{1}{\sqrt{z+1}}}\right],\\ 
    \end{split}
\end{equation}
where $z\equiv{\left[\mu + \kappa(\omega)\right]}/{\Delta}$.
Finally, the self-consistent equation simplifies to
\begin{equation}
    \frac{(z-1)^{3/2}}{2}\arctan{\frac{1}{\sqrt{z-1}}} + \frac{(z+1)^{3/2}}{2}\arctanh{\frac{1}{\sqrt{z+1}}} = z + \frac{\theta\Delta}{5k_{\mathrm{eff}}(\omega)}\left[1 + \omega^2G(\omega)\right],
\end{equation}
which is easy to solve numerically.

\end{document}